\documentclass[11pt,a4paper]{article}
\usepackage[left=1in, right=1in, top=1in, bottom=1in]{geometry}
\usepackage[toc,title,page]{appendix}
\usepackage{graphicx}
\usepackage{bm}
\newcommand*{\eqref}[1]{(\ref{#1})}
\usepackage[skip=8pt,font=scriptsize]{caption}
\usepackage{epstopdf}
\begin{document}
\begin{center}
{\bf Hydrogen energy-levels shifts induced by the atom motion. Crossover from the Lamb shifts to the motion-induced shifts.}

\vspace{.5cm}
A.I. Agafonov
\vspace{.5cm}

National Research Centre "Kurchatov Institute",

Moscow 123182, Russia

Agafonov\_AIV@nrcki.ru
\end{center}
\begin{abstract} 
When the hydrogen atom moves, the proton current generates a magnetic field which interacts with the hydrogen electron. A simple analyze shows that this interaction between the hydrogen momentum and the electron is of order of $\alpha^3\frac{v}{c}m$, where 
$\alpha$ is the fine structure constant, $v$ is the atom velocity, $c$ is the speed of light, and $m$ is the electron mass. Using the Bethe-Salpeter equation, the two velocity-dependent operators of this order are derived for the hydrogen velocity 
$\frac{v}{c}<<\alpha$. As well known, the degeneracy of the energy levels with the same principal quantum number, $n$, and the same quantum number of the total angular momentum, $j$, but the different orbital angular momenta $l=j\pm 1/2$ is removed by the radiative corrections (the Lamb shift) that are proportional to $\alpha^{5}m$. It is shown that the velocity-dependent perturbation interactions remove this degeneracy as well. There is, however, an important difference between the Lamb shifts and the energy-levels shifts induced by the atom motion. The Lamb shift is the diagonal correction to the energy separately for each of the degenerate states. The the velocity-dependent perturbation interactions result in the off-diagonal energy corrections between the mutually degenerate states. The joint effect of these two perturbations which are essentially different in their origin, is analyzed. Given their order of magnitude, the crossover from the Lamb shifts to the motion-induced shifts should occur
at the atom velocity $v=k\alpha^2 c$, where $k>>1$ is a numerical factor depended on $n$ and $j$.  An experiment used the orbital motion of the Earth, is proposed to test the theory developed. 
\end{abstract} 

\vspace{.5cm}
\vspace{.5cm}


\section{Introduction}	 
In non-relativistic quantum mechanics the energy levels of atoms and, in particular, of the hydrogen atom, do not depend on their velocities \cite{Lan}. This is due to the fact that in the non-relativistic approximation it is possible to introduce the radius vector of the center of mass. Then, in the Hamiltonian for the hydrogen atom the relative motion of the electron and proton separates from the motion of the center of mass. As a result, the degenerate energy levels of the hydrogen atom  which are inversely proportional to the square of the principal quantum number, do not depend on the motion of the atom center of mass. 
\par In quantum electrodynamic the bound-state problem for two interacting fermions has been investigated in the well-known work \cite{Bet}. The two-particle system was considered to be moving. Using the relativistic S-matrix formalism of Feynman, the relativistic bound-state equation has been derived. The following approach was used to solve this equation. Noting the difficulties in determining the coordinates of the center of mass of the system, it was selected some coordinate, more or less arbitrarily, to represent the absolute position, in time and space, of the system, 
$X_{\mu}=\beta x_{\mu 1}+(1-\beta) x_{\mu 2}$ with $\beta$ any arbitrary constant, and the relative coordinate 
$x_{\mu}=x_{\mu 1}-x_{\mu 2}$. Then, the solutions of the equation was sought for the special form of the wave function 
(see Eq. (13) \cite{Bet}),
\begin{displaymath} 
\psi(1,2)=\exp{(iK_{\mu}X_{\mu})}\psi(x_{\mu}),
\end{displaymath} 
where $K_{\mu}$ is an arbitrary constant four-dimensional vector represented the momentum and energy of the moving system.
\par Note  that this wave function $\psi(x_{\mu})$ describing the relative motion of the two particles, was assumed to be independent on  the atom four momentum $K_{\mu}$. As noted in \cite{Bet}, although the equations for $\psi(x_{\mu})$ are different in non-relativistic quantum mechanics and QED, the situation is the same in principle, namely, the integral equation 
(see Eq. (16) \cite{Bet}) for the momentum-space wave function $\psi(p_{\mu})$ depends on only four variables that are  the components of the relative momentum of the two interacting particles. It differs from the ordinary Schrodinger momentum wave function by the appearance of the "relative energy" $-p_{4}$ as a fourth independent variable.
\par The foundations of the work \cite{Bet} were then used to study many-electron systems, nuclei and quarks 
\cite{Sal,Kar,Ful,Kab,Ara,Nak,Dou,Gro,Mac,Luc,Moh,Eid,Gut}.
\par Meanwhile, in the relativistic problem the discrete atomic energies can depend on the momentum of the atom center of mass.  To demonstrate it, we present here a non-rigorous analysis, which allows us to understand the origin of some perturbation 
induced by the finite velocity of the atom. Note that this analysis is close to qualitative consideration of the spin-orbit interaction in atoms.   
\par Let the atom be moving with the constant velocity $\bf v$. Both the electron and proton have two velocity components. The first components are related to the motion of the atom center of mass, are the same for both the particles and equal to $\bf v$.
The second components are due to the relative motion of the particles.
\par The moving proton with the velocity $\bf v$ creates both the electric field $\mathbf{E}=\frac{e}{r^2}\frac{\bf r}{r}$ and the magnetic field $\mathbf{H}=\frac{e}{c}[{\bf v},\mathbf{E}]$, where $e$ is the proton charge, $\bf r$ is the relative radius vector between the electron and proton, and $c$ is the speed of light in vacuum. Here, neglecting  the terms of the order 
of $v^2/c^2$, the Lorentz factor equal to 1 is assumed. Then the magnetic field is: 
\begin{displaymath} 
\mathbf{H}=\frac{e}{c}[{\bf v}, \frac{\bf r}{r^3}]=[\nabla,\frac{{\bf v}}{c}]\Bigl(\frac{e}{r}\Bigr)=[\nabla, \mathbf{A}].
\end{displaymath} 
Hence the vector potential is 
\begin{equation}\label{1} 
\mathbf{A}=\frac{{\bf v}}{c}\Bigl(\frac{e}{r}\Bigr).
\end{equation} 
\par This expression for the vector potential can be obtained by another way. The moving proton creates the current density
${\bf j}=e{\bf v}\delta ({\bf r}_{1}-{\bf v}t)$. Then, the vector potential is
\begin{displaymath} 
\mathbf{A}({\bf r})=\frac{{e{\bf v}}}{c}\int \frac{\delta ({\bf r}_{1}-{\bf v}t)d{\bf r}_{1}}{r}, 
\end{displaymath} 
where we take into account the move of the electron with the same velocity and, hence, the relative radius vector between the electron and proton does not depend on $\bf v$. The integration over ${\bf r}_{1}$ leads to the same expression for the vector potential.
\par Substituting \eqref{1} in the Dirac Hamiltonian for the electron in the external field, $H={\bm \alpha}({\hat{\bf p}}+e\mathbf{A}) +\beta m-e\Phi$, we obtain an additional perturbation caused by the interaction of the electron spin with the magnetic field, generated by the moving proton,:
\begin{equation}\label{2} 
{\hat{W}_{e}}=-\frac{{\bm \alpha}_{e}{\bf v}}{c}V(r).
\end{equation} 
Here the ${\bm \alpha}_{e}$ matrix is used in the standard representation, the subscript $e$ means the action of the Pauli matrices on the electron spin, and  $V(r)=-\frac{e^2}{r}$ is the Coulomb interaction between the electron and proton.
\par The similar consideration of the fields generated by the moving electron with the velocity $\bf v$ leads to  
the interaction of the proton spin with the magnetic field, created by the moving electron,:
\begin{displaymath} 
{\hat{W}_{p}}=-\frac{{\bm \alpha}_{p}{\bf v}}{c}V(r),
\end{displaymath} 
where the subscript $p$ means the action of the Pauli matrices on the proton spin.
\par The commutator $[\hat{H},{\bf r}]={\bm \alpha}$, and, hence, the operators ${\bm \alpha}_{e}$ and ${\bm \alpha}_{p}$ can be considered as the operators of the relative velocities of the electron and the proton in the bound state. Since 
the relative velocity of the proton  is 1836 times less than  that for the electron,  the interaction ${\hat{W}_{p}}$  can be neglected in comparison with the interaction \eqref{2}. 
\par The interaction \eqref{2} is proportional to ${\alpha^3}\frac{v}{c}m$, where $\alpha$ is the fine structure constant and we take into account that the electron velocity $\vert <{\bm \alpha}_{e}>\vert \propto \alpha c$. Perturbations of this order will play a key role in the present paper.
\par The origin of the other perturbation that is of the same order as \eqref{2}, can be understood from the following considerations. The Dirac Hamiltonian for the electron in the Coulomb field can be reduced to the Schr$\ddot{o}$dinger one in 
the known reasonable approximations. Replacing the electron momentum in the kinetic energy operator by its canonical momentum with the vector-potential \eqref{1}, we find the electron perturbation operator:
\begin{equation}\label{3} 
{\hat{Q}_{e}}=\frac{1}{2mc}[\hat{\bf p}, \frac{\bf v}{c}V(r)]_{+}.
\end{equation} 
The similar expression for the electron perturbation operator ${\hat{Q}_{p}}$ can be written by the replacements 
${\bf p}\to -{\bf p}$ and $m\to M$ in \eqref{3}. For the above reason ${\hat{Q}_{p}}$ can be omitted.
\par We emphasize that it would be wrong to introduce directly the particle canonical momenta with the 
vector-potential \eqref{1} in the Bethe-Salpeter equation and, as well, in the Schr$\ddot{o}$dinger equation.
The operators \eqref{2} and \eqref{3} do not have the exact forms that will be obtained below. They only give an idea of the considering effect of relativistic corrections to the hydrogen discrete levels due to the atom momentum.  
The effect under investigation should manifest itself when the solutions of the relativistic two-fermion equation are sought in the form:
\begin{displaymath} 
\psi(1,2)=\exp{(iK_{\mu}X_{\mu})}\psi(x_{\mu};{\bf K}),
\end{displaymath} 
where the relative wave function $\psi(x_{\mu};{\bf K})$ is considered to depend on the four-dimensional vector ${\bf K}$ represented the atom momentum. 
\par As well known \cite{Ber}, the spin-orbit interaction in the hydrogen atom removes the degeneracy of the energy levels inherent in the non-relativistic theory, but not completely. The energy levels with the same $n$ and $j$, but different 
$l =j\pm 1/2$, remain mutually degenerate. This remaining degeneracy is removed by the so-called radiation corrections 
(the Lamb shift) \cite{Lam,Be1,Ede,Poh,Kas}, which are not taken into account in the Dirac equation for the one-electron problem.
\par The Lamb shift for the hydrogen levels $2s_{1/2}$ and $2p_{1/2}$ is equal to $\delta E_{L}=0.41\alpha^5 mc^2$ \cite{Ber}. 
Looking ahead, we note here that the atom-momentum-perturbation operators that have forms closer to \eqref{2} and \eqref{3},
also remove this remaining energy level degeneracy with the same $n$ and $j$, but different $l=j\pm 1/2$. As will be shown below, for moving hydrogen atoms the energy shift due to these interactions turn out to be proportional to 
$\delta E_{v} \propto\alpha^{3}\frac{v}{c}mc^2$ for the same levels. For this reason, there must exist the region of the atom velocities in which 
\begin{displaymath}
\delta E_{v}>\delta E_{L}.
\end{displaymath}
In this case, the levels degeneracy considered  will mostly be removed by the perturbation being studied.
\par The aim of the work is to investigate the effect of the hydrogen atom velocity on its discrete energy levels. Our analysis is carried out for the atomic velocities much less than the velocity of light in vacuum, $v<<c$. A more detailed restriction on the atom velocities will appear below. Using the Bethe-Salpeter equation we deduce the operator of the atom-momentum-induced perturbations for the hydrogen-like atoms. Then we study the shifts of the mutually degenerated energy levels due to both the radiative corrections and the interactions induced by the atom motion. Also we provide estimations for experimentally verifications of the theory developed. 
\par Natural units ($\hbar =c=1$) will be used throughout.
\section{The Bethe-Salpeter equation for the electron-proton system}
The two-fermion system considered is the electron and the proton each of which is described by the Dirac theory.
Therefore, we must use the free particle propagator, which follows from the Dirac equation for the free fermion.
Considering this case, Feynman proposed the free fermion propagator that can be written as \cite{Fey}:
\begin{equation}\label{4} 
K_{0}^{e}(2,1)=\sum_{\bf p}\frac{1}{2\varepsilon_{p}}\left[\Lambda_{e}^{+}e^{-i\varepsilon_{p}(t_{2}-t_{1})}+
\Lambda_{e}^{-}e^{i\varepsilon_{p}(t_{2}-t_{1})}\right]e^{i{\bf p}({\bf r}_{2}-{\bf r}_{1})}\theta(t_{2}-t_{1})
\end{equation} 
for the electron and
\begin{equation}\label{5} 
K_{0}^{p}(4,3)=\sum_{\bf q}\frac{1}{2\omega_{q}}\left[\Lambda_{p}^{+}e^{-i\omega_{q}(t_{4}-t_{3})}+
\Lambda_{p}^{-}e^{i\omega_{q}(t_{4}-t_{3})}\right]e^{i{\bf q}({\bf r}_{4}-{\bf r}_{3})}\theta(t_{4}-t_{3})
\end{equation} 
for the proton. Here we use following notations. For the electron propagator \eqref{4}, 
\begin{equation}\label{6} 
\Lambda_{e}^{+}=\varepsilon_{p}+m\beta_{e}+{\bm \alpha}_{e}{\bf p}, 
\Lambda_{e}^{-}=\varepsilon_{p}-m\beta_{e}-{\bm \alpha}_{e}{\bf p}, 
\end{equation} 
and $\varepsilon_{p}^{2}=m^{2}+p^{2}$, $m$ is the electron mass, $\beta_{e}$ and ${\bm \alpha}_{e}$ is the matrices in the standard representation, $\beta_{e}=\left(\begin{array}{cc} {1} & 0\\0 & {-1} \end{array}\right)$, 
${\bm \alpha}_{e}=\left(\begin{array}{cc} {0} & {\bm \sigma_{e}}\\{{\bm \sigma}_{e}} & {0} \end{array}\right)$, ${\bm \sigma}$ is
the Pauli matrices. The subscript $e$ of the Pauli matrices means their action on the electron spin. 
\par For the proton propagator \eqref{5}, 
\begin{equation}\label{7} 
\Lambda_{p}^{+}=\omega_{q}+M\beta_{p}+{\bm \alpha}_{p}{\bf q}, 
\Lambda_{p}^{-}=\omega_{q}-M\beta_{p}-{\bm \alpha}_{p}{\bf q}. 
\end{equation} 
Here $\omega_{q}^{2}=M^{2}+q^{2}$, $M$ is the proton mass, $\beta_{p}=\left(\begin{array}{cc} {1} & 0\\0 & {-1} \end{array}\right)$, ${\bm \alpha}_{p}=\left(\begin{array}{cc} {0} & {\bm \sigma_{p}}\\{{\bm \sigma}_{p}} & {0} \end{array}\right)$. 
The lower index $p$ of the Pauli matrices means their action on the proton spin. 
\par In the ladder approximation their reaction can be presented by the interaction function \cite{Fey}:
\begin{equation}\label{8} 
G^{(1)} (3,4;5,6)=-e^{2}(1-{\bm \alpha}_{e}{\bm \alpha}_{p})\delta_{+}(s_{56}^{2})\delta (3,5)\delta (4,6).
\end{equation} 
Here $s_{56}$ is the invariant distance between these particles, and $\delta_{+}(s_{56}^{2})$ is the virtual photon propagation function,
\begin{equation}\label{9} 
\delta_{+}(s_{56}^{2})=\frac{1}{4\pi^{3}}\int \frac{e^{i{\bf k}({\bf r}_{5}-{\bf r}_{6})-ik_{0}(t_{5}-t_{6})}}
{{\bf k}^2-k_{0}^{2}-i\delta}d^{4}k.
\end{equation} 
\par In the ladder approximation the Bethe-Salpeter equation for the electron-proton system is:
\begin{equation}\label{10} 
\psi (1,2)=-i\int \int \int \int d\tau_{3} d\tau_{4} d\tau_{5} d\tau_{6} K_{0}^{e}(1,3)K_{0}^{p}(2,4)G^{(1)}(3,4;5,6)\psi(5,6),
\end{equation} 
where $d\tau_{i}=d{\bf r}_{i}dt_{i}$.
\section{Separation of the atom total momentum}
As known \cite{Ber}, the Dirac equation for the electron in the external Coulomb field gives the correct values of the binding energy up to $\alpha^4$. The terms which are proportional to $\alpha^{n}$ with $n>4$, should be analyzed together with 
perturbations which are not contained in the Dirac equation. In Eq. \eqref{10} the effects of both the propagation function of the virtual photon that leads to retardation interaction between the particles, and the interaction through the vector potential created by the relative motion of the particles in the hydrogen atom, result in the energy corrections of order of $\alpha^{n}$ with $n\geq 5$ \cite{Ber}. In our consideration both these effects do not make sense to hold in this equation because the level shifts due to the proposed interaction is of order of $\alpha^{3}$, as was noted in Introduction (see \eqref{2} and 
\eqref{3}). Then, using \eqref{4}-\eqref{9}, Eq. \eqref{10} is reduced to the form:
\begin{displaymath}
\psi(1,2)=ie^{2}\int d{\bf r}_{3} \int dt_{3} \int d{\bf r}_{4}\int dt_{4} \sum_{\bf pq}\frac{1}{4\varepsilon_{p}\omega_{q}}
e^{i{\bf p}({\bf r}_{1}-{\bf r}_{3})+i{\bf q}({\bf r}_{2}-{\bf r}_{4})}
\end{displaymath}
\begin{equation}\label{11} 
\left[\Lambda_{e}^{+}e^{-i\varepsilon_{p}(t_{1}-t_{3})}+\Lambda_{e}^{-}e^{i\varepsilon_{p}(t_{1}-t_{3})}\right]
\left[\Lambda_{p}^{+}e^{-i\omega_{q}(t_{2}-t_{4})}+\Lambda_{p}^{-}e^{i\omega_{q}(t_{2}-t_{4})}\right]
\frac{1}{\vert {\bf r}_{3}-{\bf r}_{4}\vert}\psi(3,4).
\end{equation} 
\par In \eqref{11} it is convenient to separate the motion of the atom center of mass from the relative motion of the electron and proton. In relativity theory the coordinate of the center of mass cannot be defined, and the absolute position, in time and space, of the system considered, can be chosen quite arbitrarily \cite{Bet}. We consider the velocities of atoms which are much smaller than the speed of light in vacuum. Taking into account the non-relativistic limit, the choice is obvious:
\begin{equation}\label{12} 
{\bf R}=(m{\bf r}_{1}+M{\bf r}_{2})/(m+M), {\bf r}={\bf r}_{1}-{\bf r}_{2}, T=t_{1}=t_{2}. 
\end{equation} 
\par In \eqref{12} the first part is the absolute position of the atom, the second one describes the relative motion of the electron and proton, and latter, that is the equal time approach, is due to the fact that the retardation effect is absent in \eqref{11}. 
\par The solutions of Eq. \eqref{11} for the wave function are sought in the form:  
\begin{equation}\label{13} 
\psi (1,2)=\psi ({\bf r};{\bf g})e^{i{\bf gR}-iET},
\end{equation} 
where $\bf g$ is the momentum of the atom, $E$ is the total energy of the moving system. 
\par Changing to new variables of integration \eqref{12} and using \eqref{13}, Eq. \eqref{11} is rewritten as:
\begin{displaymath}
\psi ({\bf r};{\bf g})e^{i{\bf gR}-iET}=ie^{2}\int d{\bf r}^{\prime} \int d{\bf R}^{\prime} 
\int_{-\infty}^{T}dT^{\prime}
\sum_{\bf pq}\frac{1}{4\varepsilon_{p}\omega_{q}}
e^{i({\bf p}+{\bf q})({\bf R}-{\bf R}^{\prime})+i\frac{M{\bf p}-m{\bf q}}{M+m}({\bf r}-{\bf r}^{\prime})}
\end{displaymath}
\begin{equation}\label{14} 
\left[\Lambda_{e}^{+}e^{-i\varepsilon_{p}(T-T^{\prime})}+\Lambda_{e}^{-}e^{i\varepsilon_{p}(T-T^{\prime})}\right]
\left[\Lambda_{p}^{+}e^{-i\omega_{q}(T-T^{\prime})}+\Lambda_{p}^{-}e^{i\omega_{q}(T-T^{\prime})}\right]
\frac{1}{r^{\prime}}\psi ({\bf r}^{\prime};{\bf g})e^{i{\bf gR}^{\prime}-iET^{\prime}}
\end{equation} 
\par Integrating over $\bf R^{\prime}$ in \eqref{14} that gives us ${\bf p}+{\bf q}={\bf g}$, and then integrating over 
$T^{\prime}$, we obtain the following integral equation: 
\begin{equation}\label{15} 
\psi ({\bf r};{\bf g})=-e^{2}\int d{\bf r}^{\prime}\sum_{\bf p}
e^{i({\bf p}-\frac{m}{M+m}{\bf g})({\bf r}-{\bf r}^{\prime})}F({\bf p},{\bf g}-{\bf p})
\frac{1}{r^{\prime}}\psi ({\bf r}^{\prime};{\bf g}), 
\end{equation}
where
\begin{displaymath}
F({\bf p},{\bf g}-{\bf p})=\frac{1}{4\varepsilon_{{\bf p}}\omega_{{\bf g}-{\bf p}}}
\Bigl(\frac{\Lambda_{e}^{+}\Lambda_{p}^{+}}{E-\varepsilon_{p}-\omega_{{\bf g}-{\bf p}}+i\delta}+
\frac{\Lambda_{e}^{-}\Lambda_{p}^{+}}{E+\varepsilon_{p}-\omega_{{\bf g}-{\bf p}}+i\delta}+
\end{displaymath}
\begin{equation}\label{16} 
\frac{\Lambda_{e}^{+}\Lambda_{p}^{-}}{E-\varepsilon_{p}+\omega_{{\bf g}-{\bf p}}+i\delta}+
\frac{\Lambda_{e}^{-}\Lambda_{p}^{-}}{E+\varepsilon_{p}+\omega_{{\bf g}-{\bf p}}+i\delta}\Bigr)
\end{equation}
\par In \eqref{15} we introduce a new summation variable, 
\begin{equation}\label{17}      
{\bf f}={\bf p}-\frac{m}{M+m}{\bf g},
\end{equation}
and, further, we use the notations: $\varepsilon=\varepsilon_{p}$ with ${\bf p}={\bf f}+\frac{m}{M+m}{\bf g}$, and
$\omega=\omega_{\bf q}$ with ${\bf q}={\bf g}-{\bf p}=-{\bf f}+\frac{M}{M+m}{\bf g}$. Then, using \eqref{6} and \eqref{7}, the right-hand side of \eqref{16} is reduced to:
\begin{equation}\label{18}      
F=\frac{1}{2E}+\frac{2E}{\Delta}\left[ m\beta_{e}+{\bm \alpha}_{e}{\bf p}+\frac{E^2+\varepsilon^2-\omega^2}{2E}\right] 
\left[ M\beta_{p}+{\bm \alpha}_{p}{\bf q}+\frac{E^2-\varepsilon^2+\omega^2}{2E}\right]. 
\end{equation}
Here
\begin{equation}\label{19}      
\Delta=E^{4}-2E^{2}\left(\varepsilon^{2} +\omega^{2}\right)+\left(\varepsilon^{2}-\omega^{2}\right)^{2}. 
\end{equation}
\par As a result, the integral equation\eqref{15} is rewritten as:
\begin{equation}\label{20} 
\psi({\bf r};{\bf g})=-e^{2}\int d{\bf r}^{\prime}\sum_{\bf f} e^{i{\bf f}({\bf r}-{\bf r}^{\prime})}F({\bf f},{\bf g},E) 
\frac{1}{r^{\prime}}\psi ({\bf r}^{\prime};{\bf g}).
\end{equation}
\par It is easy to verify that the integral equation \eqref{20} corresponds to the following differential equation:
\begin{displaymath}
\left[E^{4}-2E^{2}\left(\varepsilon^{2} +\omega^{2}\right)+\left(\varepsilon^{2}-\omega^{2}\right)^{2}\right]
\Bigl(1+\frac{e^2}{2Er}\Bigr)\psi({\bf r};{\bf g})=
\end{displaymath}
\begin{equation}\label{21}
2E\left[m\beta_{e}+{\bm \alpha}_{e}{\bf p}+\frac{E^2+\varepsilon^2-\omega^2}{2E}\right] 
\left[M\beta_{p}+{\bm \alpha}_{p}{\bf q}+\frac{E^2-\varepsilon^2+\omega^2}{2E}\right] 
\Bigl(\frac{-e^2}{r}\Bigr)\psi({\bf r};{\bf g}).
\end{equation}
Here $\hat{\bf f}=-i{\nabla}_{\bf r}$ and 
\begin{equation}\label{22}
\varepsilon^2=m^2+(\hat{\bf f}+\frac{m}{M+m}{\bf g})^2, \omega^2=M^2+(-\hat{\bf f}+\frac{M}{M+m}{\bf g})^2
\end{equation}
\par In the equations \eqref{20} and \eqref{21} $E$ is the total energy of the moving hydrogen atom. We present $E$ as
\begin{equation}\label{23}      
E=\sqrt{(M+m+\mathcal{E}+\delta \mathcal{E}_{g})^2+g^2}+i\delta.
\end{equation}
Here $\mathcal{E}$ presents the energy levels of the hydrogen atom at $g=0$, and the value $\mathcal{E}_{g}$ is the $g$-dependent relativistic energy-levels shifts induced by the atom motion. The representation \eqref{22} is convenient since in 
non-relativistic quantum mechanics the energy levels do not depend on $g$. Further, the infinitesimal positive $\delta$ can be omitted since the bound states are investigated.  
\par The integral equation \eqref{20} and the differential equation \eqref{21} can be reduced to the Dirac equation for the hydrogen atom in the limit $m/M\to 0$, and can be transformed to the Schr$\ddot{o}$dinger Hamiltonian in the non-relativistic limit, as shown in Appendix A.  
\section{The $g$-dependent perturbation operators} 
\par The radiative corrections that are not be included in \eqref{21}, are proportional to $\alpha^5 m$. 
As was shown in the Introduction, the $g$-dependent perturbation operators is of order of $\alpha^3 \frac{v}{c}$, where
$v$ is the atom velocity, $c$ is the speed of light in vacuum. Terms of this order must be deduced from Eq. \eqref{21},
but firstly we must determine what atomic velocities are considered. To observe crossover from the Lamb shifts to the motion-induced shifts for the hydrogen atoms it turns out enough that the atom velocities
\begin{equation}\label{24}
\frac{v}{c}<<\alpha.
\end{equation}
\par Appendix B provides the corresponding transformations of the Eq. \eqref{21} in detail. As a result, it was obtained that the atom-motion-induced shift of the atomic-energy levels for hydrogen atoms is given by:
\begin{equation}\label{25} 
\delta \mathcal{E}_{g}=<\psi\vert \hat{W}+\hat{Q}\vert \psi>
\end{equation}
where the velocity dependent perturbation operators are
\begin{equation}\label{26} 
\hat{W}=\frac{{\bm \alpha}_{e}{\bf v}}{2c}V=-\frac{\alpha}{2cr} 
\left(\begin{array}{cc} {0} & {{\bf v} \bm \sigma_{e}}\\{{\bf v} {\bm \sigma}_{e}} 
& {0} \end{array}\right)
\end{equation}
and
\begin{equation}\label{27} 
\hat{Q}=\frac{{\bf v}{\hat{\bf f}}}{2mc}\Bigl(V-2\mathcal{E}\Bigr)=
i\frac{{\bf v}\nabla}{2mc}\Bigl(\frac{\alpha}{r}+2\mathcal{E}\Bigr)
\end{equation}
\par In Eq. \eqref{26} ${\bm \sigma}_{e}$ is the Pauli matrices acting on the electron bispinor. In Eq. \eqref{27} the operator 
$\hat{\bf f}$ is the momentum operator of the electron. The perturbation operators \eqref{26} and \eqref{27} exist only for the moving atoms. The represented expressions for the perturbation operators are correct only for the atom's velocities $v<<\alpha c$. 
In Eq. \eqref{25} $\psi$ is the electron wave functions of stationary states. Since $M = 1836m$ the proton mass corrections to the wave functions and the energy eigenvalues are small. Hence we can use the well-known stationary wave functions obtained for the relativistic Coulomb problem \cite{Ber}.
\par The spin-orbit interaction does not completely remove the degeneracy of the hydrogen energy levels. The levels with the same principal quantum number, $n$, and the same quantum number of the total angular momentum, $j$, but different orbital angular momenta $l=j\pm 1/2$ remain mutually degenerate with the exception of the levels with the maximal possible number $j=n-1/2$. That is, for the same $n$ the hydrogen levels $s_{1/2}$ and $p_{1/2}$, $p_{3/2}$ and $d_{3/2}$, $d_{5/2}$ and $f_{5/2}$, and so on, are mutually degenerate. This remaining degeneracy is removed by the radiative corrections (the Lamb shift), which are not taken into account in the Dirac equation. The Lamb shift for the hydrogen is proportional to $\alpha^{5} m$. In the next Section we show that the perturbation interactions \eqref{25}-\eqref{27} remove this degeneracy as well.
\section{The splitting of the hydrogen energy levels}
In the relativistic case, there are no separately conservation laws of spin and orbital angular momentum. Consequently, 
a definite value of the orbital angular momentum and its projection on any axis are absent in the stationary states.
Since there is no dedicated direction, the $x-,y-$ and $z-$axes must be equally represented. The interaction operators 
\eqref{26} and \eqref{27} are proportional to the atom velocity, $\bf v$. We choose the coordinate system such that
$v_{x}=v_{y}=v_{z}=\frac{v}{\sqrt{3}}$. Then the linear superposition of the mutually-degenerate states can be written in the following form: 
\begin{equation}\label{28} 
\psi=\sum_{m} a_{m}\psi_{nj,j-\frac{1}{2},m}+\sum_{m^{\prime}} b_{m^{\prime}}\psi_{nj,j+\frac{1}{2},m^{\prime}},
\end{equation}
where summations are taken over the secondary total angular momentum quantum numbers $m$ and $m^{\prime}$ that are 
the allowed values of the projection of the total angular momentum on the $z-$axis.
\par Our choice simplifies essentially the calculations, since in the general case, in \eqref{28} instead of the functions 
$\psi_{nj,j\pm\frac{1}{2},m}$, one should use $\frac{1}{\sqrt{3}}(\psi_{nj,j\pm\frac{1}{2},m_{x}}+\psi_{nj,j\pm\frac{1}{2},m_{y}}+
\psi_{nj,j\pm\frac{1}{2},m})$, where $m_{x}$ and $m_{y}$ are the projection of the total angular momentum on the $x-$ and 
$y-$axis. In this case ${\bf v}//z-$axis  can be chosen.
\par In the superposition \eqref{28}, for example, for the principal quantum number $n=2$ and the quantum number of the total 
angular momentum $j=1/2$ the function $\psi_{nj,j-\frac{1}{2},m}$ can be treated as the wave function of the $2s_{1/2}$ state with the orbital momentum $l=0$, and $\psi_{nj,j+\frac{1}{2},m^{\prime}}$ - as the $2p_{1/2}$ wave function with the orbital momentum $l=1$. 
\par In the standard representation, this superposition \eqref{28} is rewritten as:
\begin{equation}\label{29} 
\psi=\sum_{m} a_{m}\left(\begin{array}{cc}  
{F_{nj}^{j-\frac{1}{2}}\Omega_{j,j-\frac{1}{2},m}} 
\\ {G_{nj}^{j-\frac{1}{2}}\Omega_{j,j+\frac{1}{2},m}} \end{array}\right)+ 
\sum_{m^{\prime}} b_{m^{\prime}}\left(\begin{array}{cc} 
{F_{nj}^{j+\frac{1}{2}}\Omega_{j,j+\frac{1}{2},m^{\prime}}} 
\\ {-G_{nj}^{j+\frac{1}{2}}\Omega_{j,j-\frac{1}{2},m^{\prime}}} \end{array}\right).
\end{equation}
Here $F_{nj}^{j\pm \frac{1}{2}}(r)$, $G_{nj}^{j\pm \frac{1}{2}}(r)$ are the well-known real radial wave functions (see \cite{Ber}),
the functions $\Omega_{jlm}$ with $l=j\pm\frac{1}{2}$ are the spherical spinors which are formed by combining the spherical harmonics $Y_{l,m\pm\frac{1}{2}}$ ($m,m^{\prime}=-j,-j+1,\cdots,j-1,j$),
\begin{equation}\label{30}
\Omega_{j,j-\frac{1}{2},m}=\left(\begin{array}{cc} 
{\sqrt{\frac{j+m}{2j}}Y_{j-\frac{1}{2},m-\frac{1}{2}}} 
\\ {\sqrt{\frac{j-m}{2j}}Y_{j-\frac{1}{2},m+\frac{1}{2}}} \end{array}\right), 
\Omega_{j,j+\frac{1}{2},m}=\left(\begin{array}{cc} 
{-\sqrt{\frac{j-m+1}{2j+2}}Y_{j+\frac{1}{2},m-\frac{1}{2}}} 
\\ {\sqrt{\frac{j+m+1}{2j+2}}Y_{j+\frac{1}{2},m+\frac{1}{2}}} \end{array}\right).
\end{equation} 
\par The components of both the bispinors in \eqref{29} contain the spherical harmonics \eqref{30} of the both orders 
$l=j\pm 1/2$ that expresses the absence of a definite value of the orbital angular momentum  and its $z-$ projection in the stationary states.
\par The matrix elements of the perturbations \eqref{26} and \eqref{27} are conveniently calculated separately.
Using \eqref{29} and \eqref{30}, it is easy to see that the non-zero matrix elements of  \eqref{26} 
\begin{equation}\label{31}
W=\sum_{i=x,y,z} W_{i}=-\frac{\alpha}{2c}\int \frac{d{\bf r}}{r} \psi^{+}\Bigl[
\sum_{i=x,y,z} v_{i} \left(\begin{array}{cc} {0} & {\sigma_{i}}\\{{\sigma}_{i}} & {0} \end{array}\right)\Bigr]
\psi 
\end{equation}  
are only the off-diagonal matrix elements between the states $nl_{j}$ and $n(2j-l)_{j}$. Moreover, the matrix elements of the operator $W_{z}$ are non-zero only if the quantum numbers $m$ and $m^\prime$ are the same, $m=m^{\prime}$: 
\begin{equation}\label{32}
W_{j-\frac{1}{2},j+\frac{1}{2}}^{mm}=W_{j+\frac{1}{2},j-\frac{1}{2}}^{mm}=
\frac{\alpha}{2}\frac{v_{z}}{c}\int_{0}^{\infty}rdr \Bigl(\frac{m}{j} F_{nj}^{j-\frac{1}{2}}G_{nj}^{j+\frac{1}{2}}+
\frac{2m}{2j+2} G_{nj}^{j-\frac{1}{2}}F_{nj}^{j+\frac{1}{2}}
\Bigr).
\end{equation} 
\par The off-diagonal matrix elements of the operator $W_{x} + W_{y}$ are non-zero only if $m-m^{\prime}=\pm 1$:
\begin{displaymath}
W_{j-\frac{1}{2},j+\frac{1}{2}}^{mm^{\prime}}= \frac{\alpha}{2}\int_{0}^{\infty}rdr
\Bigl(\frac{1}{2j} F_{nj}^{j-\frac{1}{2}}G_{nj}^{j+\frac{1}{2}}+\frac{1}{2j+2} G_{nj}^{j-\frac{1}{2}}F_{nj}^{j+\frac{1}{2}}
\Bigr)
\end{displaymath}
\begin{equation}\label{33}
\left[ \frac{v_{x}-iv_{y}}{c}\sqrt{(j+m)(j-m+1)}\delta_{m^{\prime},m-1}+
\frac{v_{x}+iv_{y}}{c}\sqrt{(j-m)(j+m+1)}\delta_{m^{\prime},m+1}\right],
\end{equation} 
and
\begin{displaymath}
W_{j+\frac{1}{2},j-\frac{1}{2}}^{m^{\prime}m}=\Bigl(W_{j-\frac{1}{2},j+\frac{1}{2}}^{mm^{\prime}}\Bigr)^{*}
\end{displaymath}
\par Now consider the perturbation \eqref{27} whose matrix elements are: 
\begin{displaymath}
Q=\frac{i}{2mc}\int d{\bf r} \psi^{+}\Bigl({\bf v}\nabla \Bigl(\frac{\alpha}{r}+2\mathcal{E}\Bigr)\psi\Bigr). 
\end{displaymath}
Taking into account that the wave functions $\psi$ decrease exponentially with increasing $r$, the above expression can be reduced to a form that is more convenient for calculations:
\begin{equation}\label{34} 
Q=-\frac{i}{2mc}\int d{\bf r} \Bigl(\frac{\alpha}{r}+2\mathcal{E}\Bigr) ({\bf v}grad\psi^{+})\psi.
\end{equation}  
\par We will show that the diagonal matrix elements of the perturbation \eqref{34} vanish for the mutually-degenerate states  considered below. Avoiding cumbersome calculations, the off-diagonal matrix elements are easily obtained for each specific case of the superposition \eqref{29}-\eqref{30}.
\par Emphasize that there is an important difference between the Lamb shifts and the motion-induced shifts. The Lamb shifts lead to the diagonal corrections to the energies separately for each of the degenerate states. The interactions \eqref{26} and 
\eqref{27} depend on the velocity of the atom's motion, and result in the off-diagonal energy corrections between the mutually degenerate states. Therefore, it is of interest to analyze the joint effect of these perturbations. 
\subsection{The splitting of hydrogen $2s_{1/2}$ and $2p_{1/2}$ levels}
As well known, the radiation corrections for the hydrogen atom is
\begin{displaymath}
\delta \varepsilon_{2s}= 0.4065\alpha^5 m
\end{displaymath}  
for the $2s_{1/2}$ level and 
\begin{displaymath}
\delta \varepsilon_{2p}=-0.0050 \alpha^5 m.
\end{displaymath}  
for the $2p_{1/2}$ level. Here the two main contributions that are 1) the emission and absorption by the bound electron of virtual photons and 2) the vacuum polarization, are taken into account. The correction contributions of other effects to 
the Lamb shift will be cosidered in Section 6.
\par For the considered states the superposition \eqref{29} with account for \eqref{30} takes the form:
\begin{equation}\label{35} 
\psi=a_{\frac{1}{2}}\left(\begin{array}{cc} 
{F_{2\frac{1}{2}}^{0}Y_{0,0}} \\ {0}
\\ {-\frac{1}{\sqrt{3}}G_{2\frac{1}{2}}^{0}Y_{1,0}} \\ {\sqrt{\frac{2}{3}}G_{2\frac{1}{2}}^{0}Y_{1,1}} \end{array}\right)+ 
a_{-\frac{1}{2}}\left(\begin{array}{cc} 
{0} \\ {F_{2\frac{1}{2}}^{0}Y_{0,0}} 
\\ {-\sqrt{\frac{2}{3}}G_{2\frac{1}{2}}^{0}Y_{1,-1}} \\ {\frac{1}{\sqrt{3}}G_{2\frac{1}{2}}^{0}Y_{1,0}} \end{array}\right)+ 
b_{\frac{1}{2}}\left(\begin{array}{cc} 
{-\frac{1}{\sqrt{3}}F_{2\frac{1}{2}}^{1}Y_{1,0}} \\ {\sqrt{\frac{2}{3}}F_{2\frac{1}{2}}^{1}Y_{1,1}} \\ 
{-G_{2\frac{1}{2}}^{1}Y_{0,0}} \\ {0} \end{array}\right)+ 
b_{-\frac{1}{2}}\left(\begin{array}{cc} 
{-\sqrt{\frac{2}{3}}F_{2\frac{1}{2}}^{1}Y_{1,-1}} \\ {\frac{1}{\sqrt{3}}F_{2\frac{1}{2}}^{1}Y_{1,0}} 
\\ {0} \\ {-G_{2\frac{1}{2}}^{1}Y_{0,0}} \end{array}\right) 
\end{equation}
Here $F_{2\frac{1}{2}}^{l}(r)$, $G_{2\frac{1}{2}}^{l}(r)$ are the well-known real radial wave functions \cite{Ber},
\begin{displaymath}
\left(\begin{array}{cc} 
{F_{2\frac{1}{2}}^{l}} \\ {G_{2\frac{1}{2}}^{l}} \end{array}\right)=\frac{\pm (2\lambda)^{3/2}}{\Gamma(2\gamma+1)} 
\Big[\frac{(m\pm \varepsilon)\Gamma(2\gamma+n_{r}+1)}{4m\frac{Z\alpha m}{\lambda}\Bigl(\frac{Z\alpha m}{\lambda}-\kappa\Bigr)
n_{r}!}\Bigr]^{1/2}(2\lambda r)^{\gamma-1}e^{-\lambda r}
\end{displaymath}
\begin{equation}\label{36}
\left[ \Bigl(\frac{Z\alpha m}{\lambda}-\kappa\Bigr)F(-n_{r},2\gamma+1,2\lambda r)\mp F(1-n_{r},2\gamma+1,2\lambda r) \right],
\end{equation}  
where $F$ is the degenerate hypergeometric function, the upper signs refer to $F_{2\frac{1}{2}}^{l}$, the lower signs - to 
$G_{2\frac{1}{2}}^{l}$, the number $n_{r}=1$,  the value $\kappa=-1$ for $l=0$ and $\kappa=1$ for $l=1$. The other parameters in \eqref{36}: $Z=1$, $\gamma=\sqrt{1-\alpha^2}$, $\varepsilon/m=(1+\gamma)/\sqrt{(1+\gamma)^2+\alpha^2}$ and 
$\lambda=\sqrt{m^2-\varepsilon^2}$.
\par Denoting the state with $l=j-\frac{1}{2}=0$ by the $s$ index, and $l=j+\frac{1}{2}=1$ by the $p$ index, from \eqref{32} and \eqref{33} we obtain the off-diagonal elements of the perturbation \eqref{31}:
\begin{equation}\label{37}
W_{sp}^{mm}=W_{ps}^{mm}=(-1)^{1/2-m}\frac{v_{z}}{c} I_{1}, 
\end{equation} 
\begin{equation}\label{38}
W_{sp}^{\frac{1}{2},-\frac{1}{2}}=W_{ps}^{\frac{1}{2},-\frac{1}{2}}=\frac{v_{x}-iv_{y}}{c} I_{1}, 
\end{equation} 
\begin{equation}\label{39}
W_{sp}^{-\frac{1}{2},\frac{1}{2}}=W_{ps}^{-\frac{1}{2},\frac{1}{2}}=\frac{v_{x}+iv_{y}}{c} I_{1}. 
\end{equation} 
Here 
\begin{equation}\label{40}
I_{1}=\frac{\alpha}{2}\int_{0}^{\infty}rdr
\Bigl(F_{2\frac{1}{2}}^{0}G_{2\frac{1}{2}}^{1}+\frac{1}{3} G_{2\frac{1}{2}}^{0}F_{2\frac{1}{2}}^{1}\Bigr).
\end{equation} 
\par It remains for us to calculate the matrix elements of the perturbation \eqref{34}. It is convenient to use the spherical coordinates in which 
\begin{equation}\label{41} 
Q=-\frac{i}{2mc}\int_{0}^{\infty} \Bigl(\frac{\alpha}{r}+2\mathcal{E}\Bigr) r^2dr \int_{0}^{\pi} 
\sin \theta d\theta \int_{0}^{2\pi} d\phi 
\Bigl[\Bigl(v_{r}\frac{\partial }{\partial r} +\frac{v_{\theta}}{r}\frac{\partial }{\partial \theta} 
+\frac{v_{\phi}}{r\sin \theta}\frac{\partial }{\partial \phi} \Bigr) \psi^{+} \Bigr]\psi
\end{equation} 
and the velocity components are given by:
\begin{displaymath} 
v_{r}=v_{x} sin \theta cos \phi + v_{y} sin \theta sin \phi + v_{z} cos \theta 
\end{displaymath}
\begin{displaymath} 
v_{\theta}=v_{x} cos \theta cos \phi + v_{y} cos \theta sin \phi - v_{z} sin \theta 
\end{displaymath}
\begin{displaymath} 
v_{\phi}=-v_{x}sin \phi + v_{y}cos \phi. 
\end{displaymath}
\par For the superposition \eqref{35} all the diagonal matrix elements $Q_{ss}^{mm_{1}}$ and $Q_{pp}^{mm_{1}}$ vanish as a result of the integration over the solid angle in \eqref{41}. Calculating the off-diagonal matrix elements between the $s$ and $p$ states, one should pay attention to the following. In \eqref{35} the functions $G_{2\frac{1}{2}}^{0,1}$ are small in comparison with $F_{2\frac{1}{2}}^{0,1}$ since they have the additional multiplier $\alpha$ arising from the factor 
$(m-\varepsilon)^{1/2}$ (see \eqref{36}). Hence, taking into account the product $({\bf v}\nabla \psi^{+})\psi$ in the operator \eqref{34}, the contributions to the off-diagonal matrix elements from the functions $G_{2\frac{1}{2}}^{0,1}$  will be in 
$\alpha^2$ times smaller than that from the functions $F_{2\frac{1}{2}}^{0,1}$. Formally, instead of the functions 
$G_{2\frac{1}{2}}^{0,1}$, we can simply use zeros in the superposition \eqref{35}. As a result, we find:
\begin{equation}\label{42}
Q_{sp}^{-\frac{1}{2}-\frac{1}{2}}=-Q_{sp}^{\frac{1}{2}\frac{1}{2}}=\frac{v_{z}}{c}I_{2},
\end{equation} 
\begin{equation}\label{43}
Q_{sp}^{\frac{1}{2}-\frac{1}{2}}=\Bigl(Q_{sp}^{-\frac{1}{2}\frac{1}{2}}\Bigr)^{*}=
\frac{v_{x}-iv_{y}}{c}I_{2},
\end{equation} 
\begin{equation}\label{44}
Q_{ps}^{\frac{1}{2}\frac{1}{2}}=-Q_{ps}^{-\frac{1}{2}-\frac{1}{2}}=\frac{v_{z}}{c}I_{3}
\end{equation} 
and
\begin{equation}\label{45}
Q_{ps}^{\frac{1}{2}-\frac{1}{2}}=\Bigl(Q_{ps}^{-\frac{1}{2}\frac{1}{2}}\Bigr)^{*}=-\frac{v_{x}-iv_{y}}{c}I_{3},
\end{equation} 
where
\begin{equation}\label{46}
I_{2}=\frac{1}{6m}\int_{0}^{\infty} r^2dr
\Bigl(\frac{\alpha}{r}+2\mathcal{E}\Bigr)F_{2\frac{1}{2}}^{1}\frac{d}{dr}F_{2\frac{1}{2}}^{0} 
\end{equation} 
and
\begin{equation}\label{47}
I_{3}=\frac{1}{6m}\int_{0}^{\infty} r^2dr
\Bigl(\frac{\alpha}{r}+2\mathcal{E}\Bigr)\Big[F_{2\frac{1}{2}}^{0}\frac{d}{dr}F_{2\frac{1}{2}}^{1}+ 
\frac{2}{r}F_{2\frac{1}{2}}^{0}F_{2\frac{1}{2}}^{1}\bigr]
\end{equation} 
\par Calculating the integrals \eqref{40}, \eqref{46} and \eqref{47}  we take $\gamma=1$ instead of $\gamma=\sqrt{1-\alpha^2}$,
and $\varepsilon/m=1-\frac{\alpha^2 m}{8}$ that means neglecting the terms $\propto \alpha^n$ with $n\geq 4$. Then 
from \eqref{36} the radial wave functions are given by:
\begin{equation}\label{48}
F_{2\frac{1}{2}}^{0}=\Bigl(\frac{\alpha m}{2}\Bigr)^{3/2}e^{-\frac{\alpha mr}{2}}(2-{\alpha} mr),
\end{equation} 
\begin{equation}\label{49}
G_{2\frac{1}{2}}^{0}=-\alpha \Bigl(\frac{\alpha m}{2}\Bigr)^{3/2}e^{-\frac{\alpha mr}{2}}(1-\frac{\alpha mr}{4}),
\end{equation} 
\begin{equation}\label{50}
F_{2\frac{1}{2}}^{1}=-\frac{1}{\sqrt{3}}\Bigl(\frac{\alpha m}{2}\Bigr)^{3/2}e^{-\frac{\alpha mr}{2}}{\alpha} mr,
\end{equation} 
and
\begin{equation}\label{51}
G_{2\frac{1}{2}}^{1}=-\frac{\sqrt{3}\alpha}{2}\Bigl(\frac{\alpha m}{2}\Bigr)^{3/2}e^{-\frac{\alpha mr}{2}}(1-\frac{\alpha mr}{6}).
\end{equation} 
\par Using \eqref{48}-\eqref{51}, we obtain:
\begin{equation}\label{52}
I_{1}=-\frac{\sqrt{3}}{144}\alpha^{3}m, I_{2}=+\frac{\sqrt{3}}{144}\alpha^{3}m,
I_{3}=-\frac{\sqrt{3}}{144}\alpha^{3}m.
\end{equation} 
\par Note that the use of the exact values of $\gamma$ and $\varepsilon$ leads to corrections of the values \eqref{52} that are proportional to $\alpha^5$.
\par Taking into account of \eqref{25}, \eqref{52} and $v_{z}=\frac{v}{\sqrt{3}}$, we sum the off-diagonal matrix elements 
\eqref{37}-\eqref{39} and \eqref{42}-\eqref{45}:
\begin{displaymath}
\delta \mathcal{E}_{sp}^{\frac{1}{2}\frac{1}{2}}=W_{sp}^{\frac{1}{2}\frac{1}{2}}+Q_{sp}^{\frac{1}{2}\frac{1}{2}}=
\frac{v_{z}}{c}(I_{1}-I_{2})=-\frac{1}{72}\frac{v}{c}\alpha^{3}m
\end{displaymath}
\begin{displaymath} 
\delta \mathcal{E}_{sp}^{-\frac{1}{2}-\frac{1}{2}}=W_{sp}^{-\frac{1}{2}-\frac{1}{2}}+Q_{sp}^{-\frac{1}{2}-\frac{1}{2}}=
-\frac{v_{z}}{c}(I_{1}-I_{2})=+\frac{1}{72}\frac{v}{c}\alpha^{3}m
\end{displaymath}
\begin{displaymath}
\delta \mathcal{E}_{ps}^{\frac{1}{2}\frac{1}{2}}=W_{ps}^{\frac{1}{2}\frac{1}{2}}+Q_{ps}^{\frac{1}{2}\frac{1}{2}}=
\frac{v_{z}}{c}(I_{1}+I_{3})=-\frac{1}{72}\frac{v}{c}\alpha^{3}m
\end{displaymath}
\begin{displaymath}
\delta \mathcal{E}_{ps}^{-\frac{1}{2}-\frac{1}{2}}=W_{ps}^{-\frac{1}{2}-\frac{1}{2}}+Q_{ps}^{-\frac{1}{2}-\frac{1}{2}}=
-\frac{v_{z}}{c}(I_{1}+I_{3})=+\frac{1}{72}\frac{v}{c}\alpha^{3}m
\end{displaymath}
All the other matrix elements of the operator $\hat{W}+\hat{Q}$ are zeros.
\par Thus, we can carry out the following consideration. There are the two different perturbations for the initially degenerate states $2s_{1/2}$ and $2p_{1/2}$. Interaction between vacuum energy fluctuations and the hydrogen electron leads to the diagonal  corrections for each energy levels. These corrections named the Lamb shifts, result in the energy difference of the states 
$2s_{1/2}$ and $2p_{1/2}$. The second perturbation which is caused by the interaction between the hydrogen atom momentum and the hydrogen electron, leads to the off-diagonal matrix element between these states. Then, using \eqref{25}, the secular equation for the energies of these states has the form:  
\begin{equation}\label{53}
\det \left(\begin{array}{cccc} 
{\mathcal{E}_{g}-\delta \varepsilon_{2s}} & {0} & {\delta \mathcal{E}_{sp}^{\frac{1}{2}\frac{1}{2}}} & {0}
\\ {0} & {\mathcal{E}_{g} -\delta \varepsilon_{2s}} & {0} & {\delta \mathcal{E}_{sp}^{-\frac{1}{2}-\frac{1}{2}}}
\\ {\delta \mathcal{E}_{ps}^{\frac{1}{2}\frac{1}{2}}} & {0} & {\mathcal{E}_{g} -\delta \varepsilon_{2p}} & {0}
\\ {0} & {\delta \mathcal{E}_{ps}^{-\frac{1}{2}-\frac{1}{2}}} & {0} & {\mathcal{E}_{g} -\delta \varepsilon_{2p}}
\end{array}\right)=0.
\end{equation} 
\par Solving \eqref{53}, we find:
\begin{displaymath}
\mathcal{E}_{g\pm}=\frac{\delta \varepsilon_{2s}+\delta \varepsilon_{2p}}{2}\pm
\sqrt{\Bigl( \frac{\delta \varepsilon_{2s}-\delta \varepsilon_{2p}}{2}\Bigr)^2+\frac{1}{72^2}\alpha^6 \frac{v^2}{c^2}m^2}.
\end{displaymath}
Accordingly, the photon energy corresponded to the the transition between the $2s_{1/2}$ and $2p_{1/2}$ states is given by: 
\begin{equation}\label{54}
h\nu =\sqrt{(h\nu_{L})^2+\frac{1}{6^4}\alpha^6 \frac{v^2}{c^2}m^2}.
\end{equation} 
Here $h\nu_{L}$ is the photon energy equal to the splitting the $2s_{1/2}$ and $2p_{1/2}$ states due to the radiative shifts. Lamb and Rutherford found that $h\nu_{L}=1060$ MHz \cite{Lam}.
The second term under the root in \eqref{54} presents the splitting effect due to the interaction between the hydrogen atom momentum and the hydrogen electron.  
\par The main result is that, according to \eqref{54}, this photon energy of the transition $2s_{1/2} \to 2p_{1/2}$ is predicted to depend on the hydrogen atom velocity. The necessary discussions and evaluations will be postponed to Section 6.
\subsection{The splitting of hydrogen $3s_{1/2}$ and $3p_{1/2}$ levels}
The Lamb shift for these states is $\nu_{L}=314.894\pm0.009$MHz \cite{Bro}. 
\par For the considered states the superposition \eqref{29} with account for \eqref{30} takes the form:
\begin{equation}\label{55} 
\psi=a_{\frac{1}{2}}\left(\begin{array}{cc} 
{F_{3\frac{1}{2}}^{0}Y_{0,0}} \\ {0}
\\ {-\frac{1}{\sqrt{3}}G_{3\frac{1}{2}}^{0}Y_{1,0}} \\ {\sqrt{\frac{2}{3}}G_{3\frac{1}{2}}^{0}Y_{1,1}} \end{array}\right)+ 
a_{-\frac{1}{2}}\left(\begin{array}{cc} 
{0} \\ {F_{3\frac{1}{2}}^{0}Y_{0,0}} 
\\ {-\sqrt{\frac{2}{3}}G_{3\frac{1}{2}}^{0}Y_{1,-1}} \\ {\frac{1}{\sqrt{3}}G_{3\frac{1}{2}}^{0}Y_{1,0}} \end{array}\right)+ 
b_{\frac{1}{2}}\left(\begin{array}{cc} 
{-\frac{1}{\sqrt{3}}F_{3\frac{1}{2}}^{1}Y_{1,0}} \\ {\sqrt{\frac{2}{3}}F_{3\frac{1}{2}}^{1}Y_{1,1}} \\ 
{-G_{3\frac{1}{2}}^{1}Y_{0,0}} \\ {0} \end{array}\right)+ 
b_{-\frac{1}{2}}\left(\begin{array}{cc} 
{-\sqrt{\frac{2}{3}}F_{3\frac{1}{2}}^{1}Y_{1,-1}} \\ {\frac{1}{\sqrt{3}}F_{3\frac{1}{2}}^{1}Y_{1,0}} 
\\ {0} \\ {-G_{3\frac{1}{2}}^{1}Y_{0,0}} \end{array}\right) 
\end{equation}
Here, using \eqref{36} the radial wave functions are given by:
\begin{equation}\label{56}
F_{3\frac{1}{2}}^{0}=2\Bigl(\frac{\alpha m}{3}\Bigr)^{3/2}e^{-\frac{\alpha mr}{3}}(1-\frac{2}{3}{\alpha} mr+
\frac{2}{27}{\alpha}^2 m^2 r^2),
\end{equation} 
\begin{equation}\label{57}
G_{3\frac{1}{2}}^{0}=-\frac{\alpha}{3} \Bigl(\frac{\alpha m}{3}\Bigr)^{3/2}e^{-\frac{\alpha mr}{3}}(3-\frac{10}{9}\alpha mr
+\frac{2}{27}{\alpha}^2 m^2 r^2),
\end{equation} 
\begin{equation}\label{58}
F_{3\frac{1}{2}}^{1}=-\frac{4\sqrt{2}}{9}\Bigl(\frac{\alpha m}{3}\Bigr)^{3/2}e^{-\frac{\alpha mr}{3}}{\alpha} mr
(1-\frac{1}{6}{\alpha} mr),
\end{equation} 
and
\begin{equation}\label{59}
G_{3\frac{1}{2}}^{1}=-\frac{2\sqrt{2}\alpha}{3}{\sqrt{3}}
\Bigl(\frac{\alpha m}{3}\Bigr)^{3/2}e^{-\frac{\alpha mr}{3}}
(1-\frac{1}{3}{\alpha} mr+\frac{1}{54}{\alpha}^2 m^2 r^2).
\end{equation} 
\par Making the obvious replacement of the wave functions in \eqref{40}, \eqref{46}, and \eqref{47} by the wave functions 
\eqref{56} - \eqref{59} and calculating the corresponding integrals, we obtain:
\begin{equation}\label{60}
I_{1}=-\frac{\sqrt{2}}{18^2}\alpha^{3}m, I_{2}=+\frac{\sqrt{2}}{18^2}\alpha^{3}m,
I_{3}=-\frac{\sqrt{2}}{18^2}\alpha^{3}m.
\end{equation} 
Then the secular equation for the energies of these states takes the form \eqref{53} with the replacements 
$\delta \varepsilon_{2s} \to \delta \varepsilon_{3s}$ and $\delta \varepsilon_{2p} \to \delta \varepsilon_{3p}$.
Respectively, considering \eqref{37}-\eqref{39} and \eqref{42}-\eqref{45} but now with \eqref{60}, in \eqref{53} 
the non-zero matrix elements of the interaction \eqref{25} should be replaced by: 
\begin{displaymath}
\delta \mathcal{E}_{sp}^{\frac{1}{2}\frac{1}{2}}=W_{sp}^{\frac{1}{2}\frac{1}{2}}+Q_{sp}^{\frac{1}{2}\frac{1}{2}}=
\frac{v_{z}}{c}(I_{1}-I_{2})=-\frac{1}{2^{1/2}3^{9/2}}\frac{v}{c}\alpha^{3}m
\end{displaymath}
\begin{displaymath} 
\delta \mathcal{E}_{sp}^{-\frac{1}{2}-\frac{1}{2}}=W_{sp}^{-\frac{1}{2}-\frac{1}{2}}+Q_{sp}^{-\frac{1}{2}-\frac{1}{2}}=
-\frac{v_{z}}{c}(I_{1}-I_{2})=+\frac{1}{2^{1/2}3^{9/2}}\frac{v}{c}\alpha^{3}m
\end{displaymath}
\begin{displaymath}
\delta \mathcal{E}_{ps}^{\frac{1}{2}\frac{1}{2}}=W_{ps}^{\frac{1}{2}\frac{1}{2}}+Q_{ps}^{\frac{1}{2}\frac{1}{2}}=
\frac{v_{z}}{c}(I_{1}+I_{3})=-\frac{1}{2^{1/2}3^{9/2}}\frac{v}{c}\alpha^{3}m
\end{displaymath}
\begin{displaymath}
\delta \mathcal{E}_{ps}^{-\frac{1}{2}-\frac{1}{2}}=W_{ps}^{-\frac{1}{2}-\frac{1}{2}}+Q_{ps}^{-\frac{1}{2}-\frac{1}{2}}=
-\frac{v_{z}}{c}(I_{1}+I_{3})=+\frac{1}{2^{1/2}3^{9/2}}\frac{v}{c}\alpha^{3}m.
\end{displaymath}
\par Such modified secular equation yields the result for the photon energy of he transition between the $3s_{1/2}$ and 
$3p_{1/2}$ states: 
\begin{equation}\label{61}
h\nu =\sqrt{(h\nu_{L})^2+\frac{2}{3^9}\alpha^6 \frac{v^2}{c^2}m^2},
\end{equation} 
where the Lamb shift for these states is $\nu_{L}=314.894\pm0.009$MHz.
\par Discussion of this result is deferred to Section 6.
\subsection{The energy separation in $He^{+}$, n=2}
As was found in \cite{Lip}, the energy separation between $2s_{1/2}$ and $2p_{1/2}$ is equal to $\nu_{L}=14040.2\pm1,8$MHz 
in $He^{+}$.
\par The superposition of these states has the same form \eqref{35}. Considering $Z=2$, the radial wave functions \eqref{36}
are rewritten as:
\begin{equation}\label{62}
F_{2\frac{1}{2}}^{0}=\frac{(2\alpha m)^{3/2}}{\sqrt{2}}e^{-\alpha mr}(1-\alpha mr),
\end{equation} 
\begin{equation}\label{63}
G_{2\frac{1}{2}}^{0}=-\alpha \frac{(2\alpha m)^{3/2}}{\sqrt{2}}e^{-\alpha mr}(1-\frac{{\alpha} mr}{2}),
\end{equation} 
\begin{equation}\label{64}
F_{2\frac{1}{2}}^{1}=-\frac{(2\alpha m)^{3/2}}{\sqrt{6}}e^{-\alpha mr} \alpha mr
\end{equation} 
and
\begin{equation}\label{65}
G_{2\frac{1}{2}}^{1}=-\frac{(2\alpha m)^{3/2}}{2^{3/2}}\alpha \sqrt{6} e^{-\alpha mr}(1-\frac{{\alpha} mr}{3})
\end{equation} 
\par Substituting \eqref{62}-\eqref{65} into \eqref{40}, \eqref{46} and \eqref{47}, we obtain:  
\begin{equation}\label{66}
I_{1}=-\frac{\sqrt{3}}{36}\alpha^{3}m, 
I_{2}=+\frac{\sqrt{3}}{36}\alpha^{3}m,
I_{3}=-\frac{\sqrt{3}}{36}\alpha^{3}m.
\end{equation} 
Then the secular equation for these states in $He^{+}$ has the same form as \eqref{35} but now,  
using \eqref{37}-\eqref{39}, \eqref{42}-\eqref{45} and \eqref{66}, the nonzero elements of the matrix are given by:
\begin{displaymath}  
\delta \mathcal{E}_{sp}^{\frac{1}{2}\frac{1}{2}}=W_{sp}^{\frac{1}{2}\frac{1}{2}}+Q_{sp}^{\frac{1}{2}\frac{1}{2}}=
\frac{v_{z}}{c}(I_{1}-I_{2})=-\frac{1}{18}\frac{v}{c}\alpha^{3}m
\end{displaymath}
\begin{displaymath} 
\delta \mathcal{E}_{sp}^{-\frac{1}{2}-\frac{1}{2}}=W_{sp}^{-\frac{1}{2}-\frac{1}{2}}+Q_{sp}^{-\frac{1}{2}-\frac{1}{2}}=
-\frac{v_{z}}{c}(I_{1}-I_{2})=+\frac{1}{18}\frac{v}{c}\alpha^{3}m
\end{displaymath}
\begin{displaymath}
\delta \mathcal{E}_{ps}^{\frac{1}{2}\frac{1}{2}}=W_{ps}^{\frac{1}{2}\frac{1}{2}}+Q_{ps}^{\frac{1}{2}\frac{1}{2}}=
\frac{v_{z}}{c}(I_{1}+I_{3})=-\frac{1}{18}\frac{v}{c}\alpha^{3}m
\end{displaymath}
\begin{displaymath}
\delta \mathcal{E}_{ps}^{-\frac{1}{2}-\frac{1}{2}}=W_{ps}^{-\frac{1}{2}-\frac{1}{2}}+Q_{ps}^{-\frac{1}{2}-\frac{1}{2}}=
-\frac{v_{z}}{c}(I_{1}+I_{3})=+\frac{1}{18}\frac{v}{c}\alpha^{3}m.
\end{displaymath}
\par Solving \eqref{53} with the matrix elements presented above,  
we find the photon energy corresponded to the transition between the $2s_{1/2}$ and $2p_{1/2}$ states for the $He^{+}$ ion: 
\begin{equation}\label{67}
h\nu =\sqrt{(h\nu_{L})^2+\frac{1}{3^{4}}\alpha^6 \frac{v^2}{c^2}m^2}.
\end{equation} 
Here $\nu_{L}=14040$MHz is the separation of the energy levels $2s_{1/2}$ and $2p_{1/2}$ due to the radiative shifts. 
The second term under the root presents the splitting effect due to the interaction between the helium ion momentum and the electron of $He^{+}$.  
\section{Discussion and Conclusion}
We have established above that the remaining degeneracy of the energy levels of hydrogen-like atoms is removed not only by the well-known radiative relativistic corrections, but also by the interaction of the atomic electron with the atom momentum.
The joint effect of these two different types of the interactions is investigated. The total energy separation of the initially degenerate states for the hydrogen atom, \eqref{54} and \eqref{61}, and for the $He^{+}$ ion, \eqref{67}, was obtained.  
While the radiative corrections are proportional to $\alpha^5 m$, the shifts caused by the atom motion, are proportional to
$\alpha^3 \frac{v}{c} m$. Therefore, the crossover from the Lamb shifts to the motion-induced shifts is possible.
\par Lamb and Rutherford \cite{Lam} found experimentally that the energy separation of the hydrogen states $2s_{1/2}$ and 
$2p_{1/2}$ is equal to 1060 MHz. In these experiments the velocities of hydrogen atoms were about 8 km/s relative to the observer on the Earth. The current theoretical value of the Lamb shift is $\nu_{L}=1058.911 \pm 0.012$ MHz.
\par For the states considered, the expression \eqref{54} can be rewritten as:
\begin{equation}\label{68}
\nu=\sqrt{\nu_{L}^{2}+\nu_{H2}^{2}\frac{v^2}{c^2}},
\end{equation}
where
\begin{equation}\label{69}
\nu_{H2}=\frac{\alpha^3}{36}\frac{mc^2}{h}=1.326 THz.
\end{equation}
Both these terms under the root in \eqref{68} are equal to each other at $\frac{v_{*}}{c}=0.799\times 10^{-3}$ that means the hydrogen velocity $v_{*}=240$ km per second. Usually, stars in the Universe consist mostly of hydrogen and helium.   
In addition, stars may have significantly higher velocities than $v_{*}=240$km/s. Therefore, for such fast stars the photon energy corresponding to the radiative transition between hydrogen states $2s_{1/2}$ and $2p_{1/2}$ could depend on the star 
velocity. 
\par For the hydrogen states $3s_{1/2}$ and $3p_{1/2}$ the Lamb shift is equal to $\nu_{L}=314.894\pm0.009$MHz.
The corresponding expression \eqref{61} gives us:
\begin{equation}\label{70}
\nu=\sqrt{\nu_{L}^{2}+\nu_{H3}^{2}\frac{v^2}{c^2}},
\end{equation}
where
\begin{equation}\label{71}
\nu_{H3}=\frac{\sqrt{2}\alpha^3}{81\sqrt{3}}\frac{mc^2}{h}=0.481 THz.
\end{equation}
Both these terms under the root in \eqref{70} are equal to each other at $\frac{v_{*}}{c}=0.655\times 10^{-3}$ that is at the hydrogen velocity $v_{*}=197$ km per second.
\par Finally, we made the same estimation for the $He^{+}$ ion with $\nu_{L}=14040$MHz for the separation of the 
$2s_{1/2}$ and $2p_{1/2}$ states. The equation \eqref{67} takes the form: 
\begin{equation}\label{72}
\nu=\sqrt{\nu_{L}^{2}+\nu_{He2}^{2}\frac{v^2}{c^2}},
\end{equation}
where
\begin{equation}\label{73}
\nu_{He2}=\frac{\alpha^3}{9}\frac{mc^2}{h}=5.304 THz.
\end{equation}
Then, from \eqref{72} we obtain the hydrogen atom velocity at which the radiative shift is equal to the energy-level shift induced by the $He^{+}$ motion: $v_{*}=794$ km per second.
\par Note that in \eqref{68}, \eqref{70} and \eqref{72} the velocity $v$ is the absolute velocity of the hydrogen and hydrogen-like atoms that should be considered as a part of objects moving in the Universe. 
Let hydrogen atom be at rest relative to the observer on the Earth. The Earth has the rotation velocity $\simeq 0.5$km/s, the orbital velocity $V_{orb}\simeq 30$km/s of motion around the Sun. The Sun itself is moving with respect of the average of the stars
in its vicinity. The Sun velocity with respect to the observer's local standard of rest is about $V_{LSR}=16.5$ km/s \cite{Bin}.
\par We carry out estimations for the most studied transition $2s_{1/2} \to 2p_{1/2}$ of the hydrogen atoms. Assume that the orbital velocity $V_{orb}$ is perpendicular to the Sun velocity $V_{LSR}$. We use the experimental result obtained by Lamb and Rutherford \cite{Lam} that the energy separation between the $2s_{1/2}$ and $2p_{1/2}$ hydrogen states is equal to 
$\nu_{exp}=1060$ MHz. In their experimental setup the velocities of hydrogen atoms were about 8 km/s relative to the observer on the Earth. But unfortunately we can not take it into account, since the direction of this velocity is not known. Then, using Eq. \eqref{68}, the theoretical value of the energy separation of the $2s_{1/2}$ and $2p_{1/2}$ in the resting hydrogen atom could be: 
\begin{equation}\label{74}
\nu_{L}=\sqrt{\nu_{exp}^{2}-\delta\nu^2},
\end{equation}
where $\delta\nu$ is the atom-motion-induced correction,:
\begin{equation}\label{75} 
\delta\nu=\nu_{H2}\frac{\sqrt{V_{orb}^2+V_{LSR}^2}}{c}.
\end{equation}
Using \eqref{69}, we obtain $\nu_{L}=\nu_{exp}(1-0.01019)=1049.14$MHz. The latter is very close to the value \cite{Ber} 
\begin{equation}\label{76} 
E_{201/2}-E_{211/2}=0.41\alpha^5mc^2=1050MHz
\end{equation}
that was obtained with account for the two main radiative effects that are of order of $\alpha^5m$: 1) the emission and absorption by the bound electron of virtual photons, which leads to the change in the effective mass  and the appearance of the anomalous magnetic moment of the electron; 2) the possibility of virtual creation and annihilation of electron-positron pairs in vacuum or, in other words, the vacuum polarization, which distorts the Coulomb potential of the nucleus at distances of the order of the Compton electron wavelength. Both these contributions result in 1050,556 MHz according to data for 1996. However, Eq. \eqref{76} do not include other more subtle corrections of order of $\alpha^6m$ \cite{Bro,Ees}  which are, in total, 
$\simeq 7.355$MHz. 
\par In order to test the theory developed, the following experiment can be proposed. Let there be an  atomic hydrogen beam with the atom velocity equal to the orbital velocity of the Earth, $v_{b}=V_{orb}=30$km/s. This velocity is 3.75 times more than that in the experiments \cite{Lam}. In the first experimental geometry the beam velocity and the orbital velocity of the Earth have the same direction. In this case the frequency which corresponds to the energy separation of the $2s_{1/2}$ and $2p_{1/2}$, is denoted by $\nu_{+}$. In the second one, when the beam velocity and the orbital velocity of the Earth are opposite directed, the frequency is $\nu_{-}$. Then from \eqref{68} we find, with good accuracy, the frequency difference 
\begin{equation}\label{77} 
\delta \nu = \nu_{+}-\nu_{-}=\frac{\nu_{H2}^2}{2\nu_{L}}\frac{4V_{orb}^2}{c^2}   
\end{equation}
Using \eqref{69}, Eq. \eqref{77} gives us $\delta \nu=33.18$MHz which can be  measured experimentally.
\begin{appendices} 
\section{Test of the equations obtained} 
\renewcommand{\theequation}{A\arabic{section}}
\setcounter{section}{1}
The integral equation \eqref{20} and the differential equation \eqref{21} can be reduced to the Dirac equation for the hydrogen atom. To do this, we should put ${\bf g}$=0, ${\bf q}=0$, $\omega=M \to \infty$, and $\frac{1}{2E}\to 0$. The notation \eqref{23} is replaced as $E=M+m+\mathcal{E}$, and the equation \eqref{19} is rewritten as 
$\Delta=4M^{2}((m+\mathcal{E})^2-\varepsilon^{2})$. Accordingly, 
\begin{displaymath}
M\beta_{p}+{\bm \alpha}_{p}{\hat{\bf q}}+\frac{E^2-\varepsilon^2+\omega^2}{2E} \to 
2M \left(\begin{array}{cc} {1} & {0} \\ {0} & {0} \end{array}\right)_{p} 
\end{displaymath} 
and 
\begin{displaymath}
m\beta_{e}+{\bm \alpha}_{e}{\hat{\bf p}}+\frac{E^2+\varepsilon^2-\omega^2}{2E} \to
m\beta_{e}+{\bm \alpha}_{e}{\hat{\bf p}}+(m+\mathcal{E}).
\end{displaymath} 
\par As a result, Eq. \eqref{20} is reduced to the form:
\begin{displaymath} 
\psi({\bf r})=\int d{\bf r}^{\prime}\sum_{\bf f} e^{i{\bf f}({\bf r}-{\bf r}^{\prime})}
\frac{(m+\mathcal{E})+m\beta_{e}+{\bm \alpha}_{e}{\hat{\bf p}}}{(m+\mathcal{E})^2-\varepsilon^{2}}
V(r^{\prime})\psi ({\bf r}^{\prime}),
\end{displaymath} 
and the corresponding differential equation is 
\begin{equation}\label{A1}
\Bigl((m+\mathcal{E})^2-\varepsilon^{2}\Bigr)\psi({\bf r})=\Bigl((m+\mathcal{E})+m\beta_{e}+{\bm \alpha}_{e}
{\hat{\bf p}}\Bigr)V(r)\psi ({\bf r}),
\end{equation}
where $\mathcal{E}$ is the energy of the discrete levels, $V(r)=\frac{-e^{2}}{r}$. In the same limits Eq. (A1) can be deduced from the differential equation \eqref{21}.
\par Eq. \eqref{A1} is the Dirac equation, $({\bm \alpha}_{e}{\hat{\bf p}}+m\beta_{e}+V(r))\psi=(m+\mathcal{E})\psi$, which describes the motion of the electron in the external Coulomb field.
\par Now consider the non-relativistic limit of \eqref{21}. We assume that the ratio of atom velocity to the speed of light
$\frac{v}{c}\simeq\frac{g}{(M+m)c}<\alpha$, and take into account that $\mathcal{E}\propto \alpha^2 m$, $<V>\propto \alpha^2 m$
and $f\propto \alpha m$. Therefore this equation can be analyzed in the lowest order in the binding energy $\mathcal{E}$.
\par From Eq. \eqref{23} we have $E^2=E_{g}^2+2(M+m)\mathcal{E}$, where $E_{g}=\sqrt{(M+m)^2+g^2}$. Substituting the latter in  
Eq. \eqref{19}, the terms with $E_{g}^4$ disappear, and in the first order in the binding energy we get:
\begin{displaymath}
\Delta=8mM(M+m)\mathcal{E}-4(M+m)^2\hat{\bf f}^2.
\end{displaymath} 
\par In the first order over $\alpha^2$ the left-hand side of Eq. \eqref{21} is reduced to:
\begin{displaymath}
8(M+m)mM\left(\begin{array}{cc} {1} & {0} \\ {0} & {0} \end{array}\right)_{p} 
\left(\begin{array}{cc} {1} & {0} \\ {0} & {0} \end{array}\right)_{e} 
\Bigl(\frac{-e^2}{r}\Bigr)\psi({\bf r}).
\end{displaymath} 
Thus in the considered case Eq. \eqref{21} takes the form of the Schr$\ddot{o}$dinger equation:
\begin{displaymath}
\Bigl(\frac{{\hat{\bf f}}^2}{2\mu}-\frac{e^2}{r}\Bigr)\phi=\mathcal{E}\phi,
\end{displaymath} 
where $\mu=mM/(M+m)$ is the reduced mass, and $\phi$ is the upper spinor of the bispinor $\psi({\bf r})$.
In the same limits the Schr$\ddot{o}$dinger equation can be deduced from the integral equation \eqref{20} as well.
\section{Derivation of the equations (25), (26) and (27)} 
\renewcommand{\theequation}{B\arabic{section}}
\setcounter{section}{1}
Analyzing the equation \eqref{21}, we must take into account the following. The atom momentum $g\simeq (M+m)v$ with, according to
\eqref{24}, $v/c<<\alpha$. Considering $\mathcal{E}\propto \alpha^2 m$, $\delta \mathcal{E}_{g}\propto \alpha^3 v/c m$ and 
$f\propto \alpha m$, all terms that do not exceed $\alpha^6$ and $\alpha^5 v/c$, can be neglected. So, the terms 
$\delta \mathcal{E}_{g}^2$, $\mathcal{E}\delta \mathcal{E}_{g}$, $\mathcal{E}^2 f^2$, $\mathcal{E}^2 g^2$, $\mathcal{E}^2 fg$, and $\mathcal{E}^3$ are omitted.  
\par Eq. \eqref{21} is presented in the form:
\begin{equation}\label{B1} 
\frac{1}{4mM}\frac{\hat{\Delta}}{2E}\Bigl(1+\frac{\alpha}{2rE}\Bigr)\psi=\frac{1}{4mM}R_{e}R_{p}\Bigl(-\frac{\alpha}{r}\Bigr)\psi.
\end{equation}
Here $\hat{\Delta}$ is given by Eq. \eqref{19}. Considering \eqref{22}, \eqref{23} and \eqref{24}, the operator \eqref{19} is rewritten as:
\begin{displaymath}
{\hat{\Delta}}=4\mathcal{E}^2\Bigl[(M+m)^2+mM\Bigr]+
8\mathcal{E}\Bigl[mM(M+m)+\frac{mMg^2}{M+m}-(M+m){\hat{\bf f}}^2+(M-m){\bf g}{\hat{\bf f}}\Bigr]
\end{displaymath}
\setcounter{section}{2}
\begin{equation}\label{B2} 
+8\delta \mathcal{E}_{g}mM(M+m)-4(M+m)^2{\hat{\bf f}}^2-4g^2{\hat{\bf f}}^2+4\Bigl({\bf g}{\hat{\bf f}}\Bigr)^2.
\end{equation}
\par According to \eqref{B2} and \eqref{23}, the expansion of $(2E)^{-1}$ in the left-hand side of \eqref{B1} 
should be restricted by
\setcounter{section}{3}
\begin{equation}\label{B3}
\frac{1}{2E}=\frac{1}{2(M+m)}\Bigl(1-\frac{\mathcal{E}+\delta \mathcal{E}_{g}}{M+m}-\frac{g^2}{2(M+m)^2}\Bigr)
\end{equation}
\par Combining Eqs. \eqref{B2} and \eqref{B3}, the left-hand side of Eq. \eqref{B1} is reduced to:
\setcounter{section}{4}
\begin{displaymath}
\frac{1}{4mM}\frac{\hat{\Delta}}{2E}\Bigl(1+\frac{\alpha}{2rE}\Bigr)\psi=
\Bigl[ \mathcal{E}^2 \frac{(M+m)^2-mM}{2mM(M+m)} +
\mathcal{E}\Bigl(1+\frac{g^2}{2(M+m)^2} -\frac{{\hat{\bf f}}^2}{2mM}+\frac{(M-m){\bf g}{\hat{\bf f}}}{mM(M+m)}\Bigr)
\end{displaymath}
\begin{equation}\label{B4}
+ \delta \mathcal{E}_{g} -\frac{M+m}{2mM}{\hat{\bf f}}^2 + \frac{2({\bf g}{\hat{\bf f}})^2-g^2{\hat{\bf f}}^2}
{4mM(M+m)}\Bigr]\Bigl(1+\frac{\alpha}{2r(M+m)}\Bigr)\psi.
\end{equation}
\par Now we turn to the right side of Eq. \eqref{21}. In the representation of this equation in the form \eqref{B1} the factor
$R_{e}$ is:
\begin{displaymath}
R_{e}=m\beta_{e}+{\bm \alpha}_{e}{\bf p}+\frac{E^2+\varepsilon^2-\omega^2}{2E}. 
\end{displaymath}
Using \eqref{22}, \eqref{23} and \eqref{24}, $R_{e}$ is rewritten as:
\setcounter{section}{5}
\begin{equation}\label{B5}
R_{e}=m\beta_{e}+{\bm \alpha}_{e}\Bigl({\bf f}+\frac{m}{M+m}{\bf g}\Bigr)
+\frac{1}{2E}\Bigl[2m(M+m)+\frac{2m}{M+m}g^2+2(M+m)(\mathcal{E}+\delta \mathcal{E}_{g})+\mathcal{E}^2+2{\bf g}{\hat{\bf f}}\Bigr],
\end{equation}
where the factor $(2E)^{-1}$ is given by \eqref{B3}. 
The potential energy that is $\propto \alpha^2$, is present on the right side of the equation \eqref{B1}. Consequently in 
the approximation presented above, from \eqref{B5} we have:
\setcounter{section}{6}
\begin{equation}\label{B6} 
R_{e}=m(\beta_{e}+I)+ {\bm \alpha}_{e}\Bigl({\bf f}+\frac{m}{M+m}{\bf g}\Bigr)+\frac{M}{M+m}\mathcal{E} +
\frac{{\bf g}{\hat{\bf f}}}{M+m}.
\end{equation}
Here $I$ is the $4\times 4$ unit matrix. 
\par The another factor in the right-hand side of \eqref{B1} is given by:
\begin{displaymath}
R_{p}=M\beta_{p}+{\bm \alpha}_{p}{\bf q}+\frac{E^2-\varepsilon^2+\omega^2}{2E}
\end{displaymath}
\noindent The similar approach gives us:
\setcounter{section}{7}
\begin{equation}\label{B7} 
R_{p}=M(\beta_{p}+I)+ {\bm \alpha}_{p}\Bigl(-{\bf f}+\frac{M}{M+m}{\bf g}\Bigr)+\frac{m}{M+m}\mathcal{E}-
\frac{{\bf g}{\hat{\bf f}}}{M+m}.
\end{equation}
\par In Eqs. \eqref{B6} and \eqref{B7}  there are two ${\bm \alpha}$ matrices in the standard representation: ${\bm \alpha}_{e}$ 
acts on the electron bispinor and ${\bm \alpha}_{p}$ acts on the proton bispinor. The operator of the particle's velocity can be found as $[{\hat{H}},\bf r]$. Using the Dirac Hamiltonian for the free particle as $H$, we find 
$[{\hat{H}},\bf r]=c{\bm \alpha}$. In the bound states the momenta of the electron and proton are proportional to $f\propto {\alpha}mc$, where ${\alpha}$ is the fine structure constant. Hence the velocities of the electron and proton are proportional to 
$\vert <{\bm \alpha}_{e}>\vert \propto \alpha$ and $\vert <{\bm \alpha}_{p}> \vert \propto \frac{m}{M}\alpha$, respectively. 
\par Combining Eqs. \eqref{B6} and \eqref{B7}, the right-hand side of Eq. \eqref{B1} is reduced to:
\setcounter{section}{8}
\begin{displaymath}
\frac{1}{4mM}R_{e}R_{p}V\psi=
\frac{1}{4mM}\Bigl[m(\beta_{e}+I)+ {\bm \alpha}_{e}{\bf f}+\frac{M}{M+m}\mathcal{E}\Bigr]
\Bigl[M(\beta_{p}+I)- {\bm \alpha}_{p}{\bf f}+\frac{m}{M+m}\mathcal{E}\Bigr]V\psi
\end{displaymath}
\begin{equation}\label{B8} 
+\frac{1}{2}\left(\begin{array}{cc} {1} & {0} \\ {0} & {0} \end{array}\right)_{p}\Bigl[\frac{{\bm \alpha}_{e}{\bf g}}{M+m}+
\frac{{\bf g}{\hat{\bf f}}}{m(M+m)}\Bigr]V\psi
+\frac{1}{2}\left(\begin{array}{cc} {1} & {0} \\ {0} & {0} \end{array}\right)_{e}\Bigl[\frac{{\bm \alpha}_{p}{\bf g}}{M+m}-
\frac{{\bf g}{\hat{\bf f}}}{M(M+m)}\Bigr]V\psi.
\end{equation}
Here $V=-\frac{\alpha}{r}$.
\par According to Eq. \eqref{23}, the mass of the hydrogen atom is $M+m+\mathcal{E}+\delta \mathcal{E}_{g}$, where
$\mathcal{E}+\delta\mathcal{E}_{g}$ is the binding energy of the atom, and $\mathcal{E}>>\delta\mathcal{E}_{g}$.
At rest ($g=0$), the atomic energy levels are determined by $\mathcal{E}$. The energy levels shifts caused by
the $g$-dependent perturbations, are given $\delta \mathcal{E}_{g}$. Considering $\delta \mathcal{E}_{g}$ as a small 
correction caused by the motion of the atom, Eq. \eqref {B1} with acccount for \eqref{B4} and \eqref{B8} is divided into two equations: the first equation defines $\mathcal{E}$ and the second one - $\delta\mathcal{E}_{g}$. As a result, 
the equation for $\mathcal{E}$ takes the form:
\setcounter{section}{9}
\begin{displaymath}
\Bigl[ \mathcal{E}^2 \frac{(M+m)^2-mM}{2mM(M+m)} + \mathcal{E}\Bigl(1 -\frac{{\hat{\bf f}}^2}{2mM}\Bigr)
-\frac{M+m}{2mM}{\hat{\bf f}}^2\Bigr]\Bigl(1+\frac{\alpha}{2r(M+m)}\Bigr)\psi=
\end{displaymath}
\begin{equation}\label{B9}
\frac{1}{4mM}\Bigl[m(\beta_{e}+I)+ {\bm \alpha}_{e}{\bf f}+\frac{M}{M+m}\mathcal{E}\Bigr]
\Bigl[M(\beta_{p}+I)- {\bm \alpha}_{p}{\bf f}+\frac{m}{M+m}\mathcal{E}\Bigr]V\psi
\end{equation}
One can see from Eq. \eqref{B9} that the binding energy $\mathcal{E}$ does not depend on the hydrogen momentum.
\par The second equation for $\delta \mathcal{E}_{g}$ is given by:
\setcounter{section}{10}
\begin{displaymath}
\Bigl[ \delta \mathcal{E}_{g}+\mathcal{E}\Bigl(\frac{g^2}{2(M+m)^2}+\frac{(M-m){\bf g}{\hat{\bf f}}}{mM(M+m)}\Bigr)
+ \frac{2({\bf g}{\hat{\bf f}})^2-g^2{\hat{\bf f}}^2}
{4mM(M+m)}\Bigr]\psi=
\end{displaymath}
\begin{equation}\label{B10}
\frac{1}{2}\left(\begin{array}{cc} {1} & {0} \\ {0} & {0} \end{array}\right)_{p}\Bigl[\frac{{\bm \alpha}_{e}{\bf g}}{M+m}+
\frac{{\bf g}{\hat{\bf f}}}{m(M+m)}\Bigr]V\psi
+\frac{1}{2}\left(\begin{array}{cc} {1} & {0} \\ {0} & {0} \end{array}\right)_{e}\Bigl[\frac{{\bm \alpha}_{p}{\bf g}}{M+m}-
\frac{{\bf g}{\hat{\bf f}}}{M(M+m)}\Bigr]V\psi.
\end{equation}
Now we analyze Eq. \eqref{B9}. It is reduced to the form:
\setcounter{section}{11}
\begin{displaymath}
\frac{m}{\mu}\Bigl[ \mathcal{E}^2\Bigl(1- \frac{\mu}{M+m}\Bigr) +2\mu \mathcal{E}\Bigl(1 -\frac{{\hat{\bf f}}^2}{2mM}\Bigr)
-{\hat{\bf f}}^2\Bigr]\Bigl(1+\frac{\alpha}{2r(M+m)}\Bigr)\psi=
\end{displaymath}
\begin{equation}\label{B11}
\Bigl[m+m\beta_{e}+{\bm \alpha}_{e}{\bf f}+\frac{\mu}{m}\mathcal{E}\Bigr]
\Bigl[\left(\begin{array}{cc} {1} & {0} \\ {0} & {0} \end{array}\right)_{p}-\frac{{\bm \alpha}_{p}{\bf f}}{2M}
+\frac{m\mathcal{E}}{2M(M+m)}\Bigr]V\psi
\end{equation}
In the left-hand side of \eqref{B11} the terms $\mathcal{E}^2\frac{\mu}{M+m}$ and $\mathcal{E}f^2\frac{\mu}{2mM}$ are proportional to $\alpha^4\frac{m}{M}$. In the left-hand side of the equation the terms $\frac{{\bm \alpha}_{p}{\bf f}}{2M}$ and 
$\frac{m\mathcal{E}}{2M(M+m)}$ $\propto \alpha^2 \frac{m^2}{M^2}$. Omitting  these four terms, \eqref{B11} is rewritten as:
\setcounter{section}{12}
\begin{equation}\label{B12}
\frac{m}{\mu}\Bigl[ (\mu+\mathcal{E})^2- \mathcal{E}_{\mu}^2 \Bigr]\Bigl(1+\frac{V}{2(M+m)}\Bigr)\psi=
\Bigl[m+m\beta_{e}+{\bm \alpha}_{e}{\bf f}+\frac{\mu}{m}\mathcal{E}\Bigr]V\psi(\bf r; g=0),
\end{equation}
where $\mu=Mm/(M+m)$ is the reduced mass, and $\varepsilon_{\mu}=\sqrt{\mu^2+{\hat{\bf f}}^2}$. 
\par It is interesting to compare Eq. \eqref{B12}, in which the finite mass of the proton is taken into account, with Eq. 
(A1), which is the Dirac equation for the electron in the external Coulomb field. It is seen that in the limit 
$M\to \infty$ the equation \eqref{B12} goes to Eq. (A1).
\par Finally we consider Eq. \eqref{B10}. In the Introduction we showed that $\delta \mathcal{E}_{g}$ must be of order of
$\alpha^3 \frac{v}{c}$. In the left-hand of the equation the first term in parentheses and the last term in square brackets are proportional to $\alpha^2 \frac{v^2}{c^2}$. Taking into account \eqref{24}, both these terms can be neglected.
The remaining term in equation, 
\begin{displaymath}
\mathcal{E}\frac{(M-m){\bf g}{\hat{\bf f}}}{mM(M+m)}\propto \alpha^3\frac{v}{c}m.
\end{displaymath}
Analyzing the right-hand side of \eqref{B10}, it should be considered that in the bound states the velocities of the electron and proton are proportional to $\vert <{\bm \alpha}_{e}>\vert \propto \alpha$ and $\vert <{\bm \alpha}_{p}> \vert \propto \frac{m}{M}\alpha$, respectively. Therefore, the second term is small in comparison with the first one. Taking into account that 
$<V>\propto \alpha^2 m$, both the remaining terms are proportional to $\alpha^3 \frac{v}{c}$. 
\par Summarizing the above, Eq. \eqref{B10} takes the form:
\setcounter{section}{13}
\begin{equation}\label{B13} 
\delta \mathcal{E}_{g}=<\psi\vert \hat{W}+\hat{Q}\vert \psi>
\end{equation}
where the perturbation operators are
\setcounter{section}{14}
\begin{equation}\label{B14} 
\hat{W}=\frac{{\bm \alpha}_{e}{\bf v}}{2c}V=-\frac{\alpha}{2cr} 
\left(\begin{array}{cc} {0} & {{\bf v} \bm \sigma_{e}}\\{{\bf v} {\bm \sigma}_{e}} 
& {0} \end{array}\right)
\end{equation}
and
\setcounter{section}{15}
\begin{equation}\label{B15} 
\hat{Q}=\frac{{\bf v}{\hat{\bf f}}}{2mc}\Bigl(V-2\mathcal{E}\Bigr)=
i\frac{{\bf v}\nabla}{2mc}\Bigl(\frac{\alpha}{r}+2\mathcal{E}\Bigr).
\end{equation}
\par In Eqs. \eqref{B13}- \eqref{B15} the wave functions $\psi$ are given. They are determined by \eqref{B12} or \eqref{A1}, since $M = 1836m$ and the proton mass corrections to the wave functions and the eigenvalues are small. 
Therefore, Eq. \eqref {B13} should be regarded as the equation for the atom-motion-induced corrections $\delta \mathcal{E}_{g}$ to the hydrogen energy levels for the known wave functions.
\end{appendices}
 
\end{document}